\documentclass{aa}
\usepackage{graphics}
\usepackage{epsfig}
\usepackage{txfonts}
\begin{document}
\title{The oxygen abundance gradient in \object{M101}: the reliability of the $\boldmath{P}$ method}
\author{Bernab\'e Cedr\'es \inst{1} \and Miguel Alejandro Urbaneja \inst{1} \and Jordi Cepa \inst{1,2}}
\institute{Instituto de Astrof\'{\i}sica de Canarias, E-38200 La Laguna, Tenerife, Spain
\and Departamento de Astrof\'{\i}sica, Facultad de F\'{\i}sica, Universidad de La Laguna, E-38071 La Laguna, Tenerife,
Spain}
\offprints{Bernab\'e Cedr\'es}
\mail{bce@ll.iac.es}
\abstract{We present the oxygen abundance determination for 90 \ion{H}{ii} regions in the inner parts of
the grand design galaxy M101. The abundances were derived employing the $P$ method (Pilyugin 2001a). A comparison 
is made with 
previous
determinations using another calibration and direct measurements of electron temperature to derive the oxygen 
abundance. The results show  agreement with the abundances derived from the electron temperature method and 
also show that the older calibration is not as accurate as the $P$ method.
}
\titlerunning{The oxygen abundance at M101}
\maketitle

\section{Introduction}
The determination of  oxygen abundance is a critical stage prior to deriving the value for the metallicity in galaxies 
and the equivalent 
abundances for several other elements, such as sulfur, nitrogen or argon. 
The preferred method for determining the oxygen abundance in galaxies using H~{\sc ii} regions is through electron
temperature-sensitive lines (the so-called $T_{\rm e}$ method), such as the [O~{\sc iii}]$\lambda$4363 or 
[\ion{O}{iii}]$\lambda$7325 
auroral lines 
 (Searle 1971; Rosa 1981; Garnett \& Kennicutt 1994; Kennicutt et al.\ 2003).
However, these lines are not always available: for oxygen-rich regions, the oxygen line [O~{\sc iii}]$\lambda$4363 is
weak and difficult to detect, so there are not many direct abundance determinations from the inner parts of galaxies.

Other methods are based on ``empirical'' calibrations of metallicity employing strong-line abundance estimators. These
methods are based on direct measurements of the electronic temperature of low metallicity regions and in theoretical
models for high metallicity regions. One method with widespread acceptance and use is the $R_{23}$-method,
suggested by Pagel et al. (1979). It is based in the oxygen line ratio, $R_{23}=([{\rm O II}]\lambda\lambda 3726,29 +
[{\rm O III}]\lambda\lambda 4959,5007)/{\rm H}\beta$. There are different calibrations using the $R_{23}$ ratio, such as
those of Dopita \& Evans (1986), Edmunds \& Pagel (1984), McCall et al. (1985), McGaugh (1991) or Zaritsky et al.\ (1994).
However, this indicator presents one great disadvantage: the derived abundances depend strongly  on the 
$R_{23}$-\ion{O}/\ion{H} calibration (Kewley \& Dopita 2002; Cedr\'es 2003).

Moreover, for \object{M101} Kennicutt et al.\ (2003)  have recently found systematic differences up to a factor 3 between abundances 
derived from 
some empirical
calibrations and those derived from the direct method, the latter being lower.

Smartt et al.\ (2001) and Trundle et al.\ (2002) have shown that in the Local Group spiral galaxy M31  oxygen abundances 
of
B supergiant atmospheres are also systematically lower than those obtained by classical $R_{23}$-\ion{O}/\ion{H}
calibrations. However, Monteverde et al.\ (2000) found very good agreement between B supergiant abundances (obtained
in a similar way as in the aforementioned references) and the abundances derived from the $T_{\rm e}$ method in the 
interstellar
medium in M33. Thus, it is clear that more comparisons are required.

New methods for  abundance determinations using strong lines have been developed recently. These methods achieve a
better approximation to the results obtained with the $T_{\rm e}$ method. One of these new calibrations, the $P$ method, is 
proposed by
Pilyugin (2001a, 2001b).

In this paper we present an estimate of the oxygen abundance for the inner H~{\sc ii} regions of M101 using the
$P$ method with data from direct imaging observations, which give us a larger number of regions when compared with
spectroscopic methods, these allowing  better sampling of the disc both spatially and in \ion{H}{ii} region luminosity, 
and with less telescope time.
\section{Data}
The  data have been taken from Cedr\'es \& Cepa (2002). These data were obtained using the direct imaging method
through narrow band filters. The observational methods and the reduction and calibration processes are described in 
Cedr\'es
\& Cepa (2002).  All the regions are extinction-corrected and the H$\beta$ line is also corrected for underlying 
absorption (Cedr\'es \& Cepa 2002).
 For the total sample of 338 regions, we selected regions with data in [\ion{O}{ii}]$\lambda\lambda$3727,3729 and 
 [\ion{O}{iii}]$\lambda\lambda$4959,5007, where we assumed that 
 [\ion{O}{iii}]$\lambda\lambda$4959,5007 = 1.34[\ion{O}{iii}]$\lambda$5007, because for our observations there were only 
 filters available for the [\ion{O}{iii}]$\lambda$5007 line. After the
determination of the value of the oxygen abundance, we rejected those regions with an error in the determination of the 
oxygen
abundance larger than 0.23 dex. These errors were determinated from the propagation of the uncertainties in the 
line strengths.
We obtained a total of 90 H~{\sc ii} regions with $R\le0.3R_0$, where $R_0$ is the disc isophotal diameter ($R_0=14.42$ 
arcmin $=$ 32.4 kpc; de Vaucouleurs et al.\ 1991). We  selected only the inner parts of the galaxy and regions with
12 + log(O/H) $\ge$ 8.4 for two reasons: to avoid the problem caused by  a systematic change in the slope of the 
gradient with strong line
determinations of the abundance (Kennicutt et al.\ 2003), and to be sure that all the regions were in the high 
metallicity regime, thus avoiding  uncertainties due to the low metallicity branch of Pilyugin's (2001a) calibration.
Our cut off is approximately 0.2 dex over the limit indicated  by Pilyugin (2001a). In Table 3 \footnote {Table 3 is only available in electronic form at the CDS via anonymous ftp to cdsarc.u-strasbg.fr (130.79.128.5)} are listed 
the results for all the regions. The first column is the region identification number from Cedr\'es \& Cepa (2002): 
the second column is the oxygen abundance.
\section{Results}
The oxygen abundance was derived using the method proposed by Pilyugin (2001a,b), using the following
expression:
\begin{equation}
12+\log ({\rm O}/{\rm H})_P=\frac{R_{23}+54.3+59.45P+7.31P^2}{6.07+6.71P+0.37P^2+0.243R_{23}},
\end{equation}
where
\begin{equation}
R_{23}=\frac{I_{[{\rm OII}]\lambda\lambda3727,29}+I_{[{\rm OIII}]\lambda\lambda4959,5007}}{H\beta}
\end{equation}
and
\begin{equation}
P=\frac{I_{[{\rm OIII}]\lambda\lambda4959,5007}}{I_{[{\rm OII}]\lambda\lambda3727,29}+I_{[{\rm
OIII}]\lambda\lambda4959,5007}}.
\end{equation}
This calibration is valid in moderately high-metallicity H~{\sc ii} regions (Pilyugin 2001a).

\begin{figure}
\epsfig{file=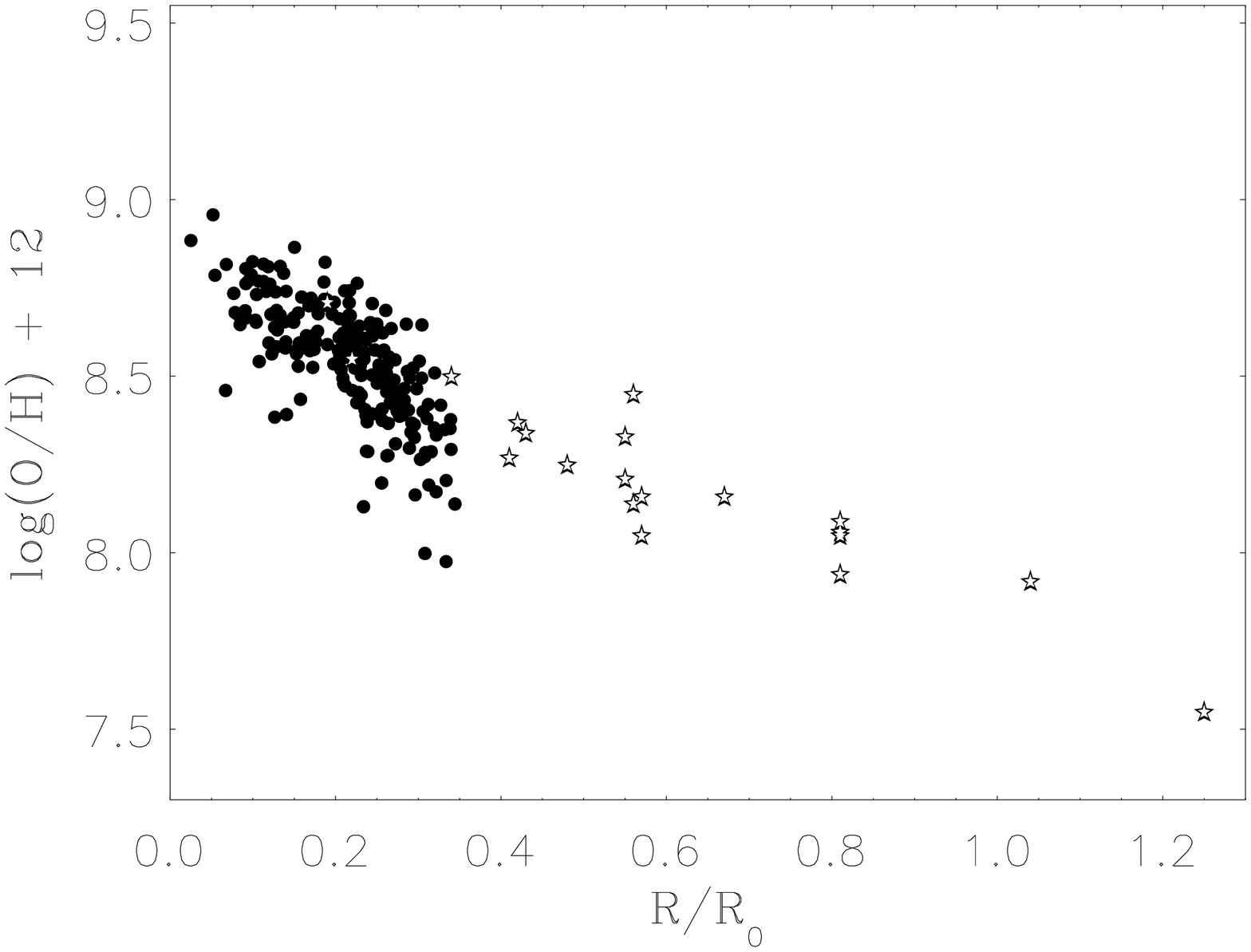,angle=90,width=88mm}
\caption{Oxygen abundance for M101 as a function of galactocentric radius. The dots represent all the regions with 
detected flux from [\ion{O}{ii}] and [\ion{O}{iii}] lines, from 
Cedr\'es
\& Cepa (2002), using the $P$ method. The stars represent the data from Kennicutt et al.\ (2003), using electron 
temperature-sensitive lines.}
\label{tod}
\end{figure}

 In Figure \ref{tod} we represent the oxygen abundance versus the galactocentric radius divided by $R_0$, for all the 
 regions with data in [\ion{O}{ii}] and [\ion{O}{iii}] from Cedr\'es \& Cepa (2002). The change in slope (compared to 
 the $T_{\rm e}$ data for $12+\log (\rm{O}/\rm{H})<8.4$) is clear in Figure \ref{tod}, as noted in Kennicutt et al.\ 
 (2003), 
 so it is not advisable to use this calibration for the outer, lower metallicity, parts of the galaxy. This behaviour is 
 also observed (but with fewer regions) for this galaxy in Pilyugin (2001b).

\begin{figure}
\epsfig{file=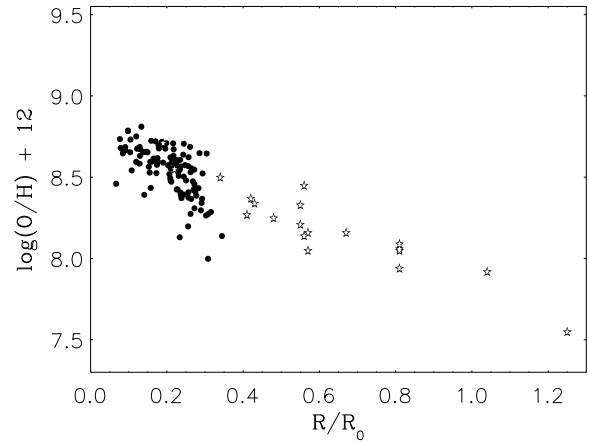,angle=90,width=88mm}
\caption{Oxygen abundance for M101 as a function of galactocentric radius. The dots represent all the regions with 
errors less than 0.23 dex, from 
Cedr\'es
\& Cepa (2002), using the $P$ method. The stars represent the data from Kennicutt et al.\ (2003), using electron 
temperature-sensitive lines.}
\label{err}
\end{figure}

 In Figure \ref{err} we now represent the oxygen abundance as a function of the galactocentric radius divided by $R_0$ 
 for all the regions with errors less than 0.23 dex. There is a good correlation between the data derived from the 
 $P$ method (dots) and the data derived through the $P$ method (stars), and the 
 tightness of the relation is now clearer.

\begin{figure}
\epsfig{file=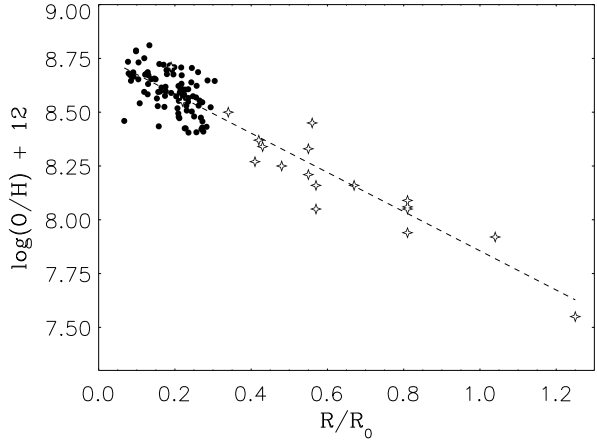,angle=90,width=88mm}
\caption{Oxygen abundance for M101 as a function of galactocentric radius. The dots represent the regions from 
Cedr\'es
\& Cepa (2002) using the $P$ method with an abundance greater than 8.4 dex and errors less than 0.23 dex. 
The stars represent the data from Kennicutt et al.\ (2003) using electron 
temperature-sensitive lines. The linear fit to the data is shown by the dashed line.}
\label{grad}
\end{figure}

In Figure \ref{grad} we represent the derived oxygen abundance as a function of the galactocentric radius divided by 
$R_0$ for the selected regions with errors less than 0.23\ dex and abundance larger than 8.4, to avoid the 
uncertainties due to the low metallicity branch of the $P$ method
 From Figure \ref{grad}, it is clear that there is 
very good agreement between our data and those of Kennicutt et al.\ (2003). Moreover, our data cover a zone of
 high
metallicity where only two regions are available from temperature-sensitive line method because of the low excitation of
 the H~{\sc ii} regions in the
inner disc of the galaxy and the stronger continuum, which make the detection of the auroral lines difficult
(Kennicutt et al.\ 2003). 

In Table \ref{anch} we present the equivalent width for the [\ion{O}{ii}] and [\ion{O}{iii}] lines for our data 
(columns two and four) and for Kennicutt et al. (2003) data (columns three and five). The equivalent widths from
 Kennicutt et al.\ (2003) were derived assuming a value for the continua equal to those measured by 
 Cedr\'es \& Cepa (2002). 
In Table \ref{coin} we present the coincidental regions. The first column is the region 
identification (from Kennicutt \& Garnett 1996), the second column is the $T_{\rm e}$-derived metallicity, the third 
column is the $P$-derived metallicity from data in Cedr\'es \& Cepa (2002) and the last column is the $P$-derived 
metallicity from Kennicutt \& Garnett (1996) data. It is clear that both regions present an agreement  within 
the error 
limits for our $P$-derived data and the $T_{\rm e}$-derived data.
\begin{table}
\caption{Equivalent widths for coincidental regions}
\begin{tabular}{c|c|c|c|c}
Region & EW([\ion{O}{ii}]) & EW([\ion{O}{ii}]) & EW([\ion{O}{iii}]) & EW([\ion{O}{iii}])\\
 & & (K2003)&  & (K2003) \\ 
\hline
H336 & 495\AA & 570\AA & 183\AA & 226\AA\\
H1013 & 1002\AA & 816\AA & 1135\AA & 1732\AA\\
\hline
\end{tabular}
\label{anch}
\end{table}
\begin{table}
\caption{Abundance derived for the coincidental \ion{H}{ii} regions}
\begin{tabular}{c|c|c|c}
Region  & $T_e$ & $P$ (This work) & $P$ (K\&G)\\
\hline
H336 & 8.55$\pm$0.16 & 8.63$\pm$0.22 & 8.59\\
H1013 & 8.71$\pm$0.05 & 8.53$\pm$0.20 & 8.60\\
\hline
\end{tabular}
\label{coin}
\end{table}
The best fit for all the data (including our strong-line derived abundance regions and the electron temperature-sensitive
lines regions from Kennicutt et al.\ 2003) is:
\begin{equation}
12+\log ({\rm O}/{\rm H})=8.767(\pm 0.021)-0.911(\pm 0.033)R/R_0,
\label{gradi}
\end{equation}
which gives us a gradient of $-0.028\pm0.01$ dex/kpc.

The fit in eq.\ (\ref{gradi}) has an abundance scatter of the same order and presents considerable agreement with that 
derived by Kennicutt et al.\ (2003) using $T_{\rm e}$-based data alone:
\begin{equation}
12+\log ({\rm O}/{\rm H})=8.76(\pm 0.06)-0.90(\pm 0.08)R/R_0.
\end{equation}
When comparing these results with those presented in Cedr\'es \& Cepa (2002) employing the empirical calibration of
Zaritsky et al. (1994), it is clear that there
is a large difference in the determination of the oxygen abundance. Taking into account the data 
from Kennicutt et al.\ (2003), we may assume that the early determinations of the metallicity of the central parts 
of the galaxy in Cedr\'es \& Cepa (2002) were not as accurate as those presented here. Moreover, the shape of the 
abundance gradient is different. From Cedr\'es \& Cepa (2002) it seems that there is a change in slope for the outer 
parts of the galaxy. However, such behaviour it is not shown here with the $T_{\rm e}$-derived data. Therefore, as 
proposed 
in Cedr\'es \& Cepa (2002), this turnover may be due to a systematical error in the  calibration employed by Zaristky 
et al. (1994).
Our results seem to indicate that the gradient is linear right through to the inner regions of the galaxy. This assumption 
is corroborated by the coincidence between the two regions with $T_{\rm e}$ data and $P$ data (H336 and H1013). However to 
fully confirm the linearity and the absolute value of the $P$ method as a tool for abundance measurement in the inner 
parts of galaxies, more determinations of abundance employing  the $T_e$ and  $p$ methods simultaneously are 
required.
\section{Conclusions}
We have obtained the oxygen abundance through the $P$ method for 90 H~{\sc ii} regions of the inner parts of M101.

There is very good agreement between our data and the derived abundances from the $T_{\rm e}$ method for regions with
high metallicity. Compared with spectroscopic methods, these results present a larger number of regions 
than any previous study. The dispersion of the data shows that, even for the larger uncertainties, the 
data are almost as reliable as those of spectroscopic studies employing the $p$ or the $T_{\rm e}$ method. 
Moreover, the direct imaging method is less time-consuming because only two observing nights are required to 
obtain data for more than a hundred regions from one galaxy.

The $P$ method has proved to be a useful tool for determining  oxygen abundances in the inner zones of 
 galaxies, where
 auroral lines are difficult to measure and the metallicity is moderately high. Moreover, direct imaging techniques 
 proved  superior when considering observing time and the number of  regions observed.

\begin{acknowledgements}
This work was supported by the Spanish Plan Nacional de Astronom\'{\i}a y Astrof\'{\i}sica under grant AYA2002-01379.
\end{acknowledgements}


\begin{thebibliography}{}
\bibitem[2003]{}
Cedr\'es, B. 2003, Ph.D. Thesis, Universidad de La Laguna
\bibitem[2002]{belito}
Cedr\'es, B., \& Cepa, J. 2002, A\&A, 391, 809
\bibitem[1991]{}
de Vaucouleurs, G., de Vaucouleurs, A., Corwin, H. G., Buta, R. J., Paturel, G., \& Fouque, P. 1991, Third Reference
Catalog of Bright Galaxies, University of Texas Press
\bibitem[1986]{}
Dopita, M. A., \& Evans, I. N. 1986, ApJ, 307, 431
\bibitem[1984]{}
Edmunds, M. G., \& Pagel, B. E. J. 1984, MNRAS, 211, 507
\bibitem[1994]{}
Garnett, D. R., \& Kennicutt, R. C. 1994, ApJ, 426, 123
\bibitem[2003]{}
Kennicutt, R. C., Bresolin, F., \& Garnett, D. R. 2003, ApJ, 591, 801
\bibitem[1996]{}
Kennicutt, R. C., Garnett, D. R. 1996, ApJ, 504, 518
\bibitem[2002]{}
Kewley, L. S., \& Dopita, M. A. 2002, ApJSS, 142, 35
\bibitem[1985]{}
McCall, M. L., Rybski, P. M., \& Shields, G. S. 1985, ApJS, 57, 1
\bibitem[1991]{}
McGaugh, S. S. 1991, ApJ, 380, 140
\bibitem[2000]{}
Monteverde, M. I., Herrero, A., \& Lennon, D. J. 2000, ApJ, 545, 813
\bibitem[1979]{}
Pagel, B. E. J., Edmunds, M. G., Blackwell, D. E., Chun, M. S., \& Smith, G. 1979, MNRAS, 189, 95
\bibitem[2001]{}
Pilyugin, L. S. 2001a, A\&A, 369, 594
\bibitem[2001]{}
Pilyugin, L. S. 2001b, A\&A, 373, 56
\bibitem[1981]{}
Rosa, M., 1981, PhD thesis, Landessternwarte Heidelberg/K\"onigstulh
\bibitem[1971]{}
Searle, L. 1971, ApJ, 168, 327
\bibitem[2001]{}
Smartt, S. J., Crowther, P. A., Dufton, P. L., Lennon, D. J., Kudritzki, R. P., Herrero, A., McCarthy, J. K., \&
Bresolin, F. 2001, MNRAS, 325, 257
\bibitem[2002]{}
Trundle, C., Dufton, P. L., Lennon, D. J., Smartt, S. J., \& Urbaneja, M. A. 2002, A\&A, 395, 519
\bibitem[1994]{}
Zaritsky, D., Kennicutt, R. C., \& Huchra, J. P. 1994, ApJ, 420, 87
\end{thebibliography}
\end{document}